\documentclass{PoS}

\usepackage{lineno}

\title{The Apollo ATCA Platform}

\ShortTitle{The Apollo ATCA Platform}

\author{A.~Albert, J.~Butler, Z.~Demiragli, K.~Finelli, D.~Gastler, \speaker{E.~Hazen}, J.~Rohlf, S.~Yuan
        \\
        Boston University\\
        E-mail: \email{hazen@bu.edu}}

\author{T.~Costa~de~Paiva, V.~Martinez~Outschoorn, S.~Willocq \\
  University of Massachusetts Amherst}

\author{C.~Strohman, P.~Wittich\\ Cornell University}

\author{R.~Glein, K.~Ulmer\\
        University of Colorado}

\abstract{We have developed a novel and generic open-source platform -
  Apollo - which simplifies the design of custom Advanced
  Telecommunications Computing Architecture (ATCA) blades by factoring
  the design into generic infrastructure and application-specific
  parts. The Apollo "Service Module" provides the required ATCA
  Intelligent Platform Management Controller, power entry and
  conditioning, a powerful system-on-module (SoM) computer, and
  flexible clock and communications infrastructure. The Apollo
  "Command Module" is customized for each application and typically
  includes two large field-programmable gate arrays, several
  hundred optical fiber interfaces operating at speeds up to 28 Gbps,
  memories, and other supporting infrastructure.  The command and
  service module boards can be operated together or independently on
  the bench without need for an ATCA shelf.}

\FullConference{Topical Workshop on Electronics for Particle Physics TWEPP2019\\
		2-6 September 2019\\
		Santiago de Compostela - Spain}

\newcommand{\iic}{$\mathrm{I^2C}$~}

\begin{document}


\section{Introduction}

The development of high-performance Advanced Telecommunications
Computing Architecture (ATCA) blades for high-energy physics applications
has proven to be quite challenging~\cite{bib:esm,bib:Serenity}. 
Many problems must be solved, including: delivery of adequate 
cooling for 400~W of power; high-performance communications
interfaces for control, monitoring and data acquisition; optical fiber
management; and industry-standard debug and programming interfaces for
routine monitoring and recovery of "bricked" modules.
The Apollo platform provides a relatively simple hardware environment
and firmware and software toolkit which may be used to develop ATCA
blades without reinventing all the required infrastructure.

We have constructed and tested 10 demonstrator modules which are
currently in operation at test stands at collaborator institutes and
at CERN.  We describe the design in detail, show some test results and
plans for further development and deployment in various LHC
experiments.

This paper describes a demonstrator of the Apollo board developed in 2018
and 2019 at Boston University and Cornell University.

\section{Overview}

\begin{figure}
  \begin{centering}
    \includegraphics[width=\textwidth]{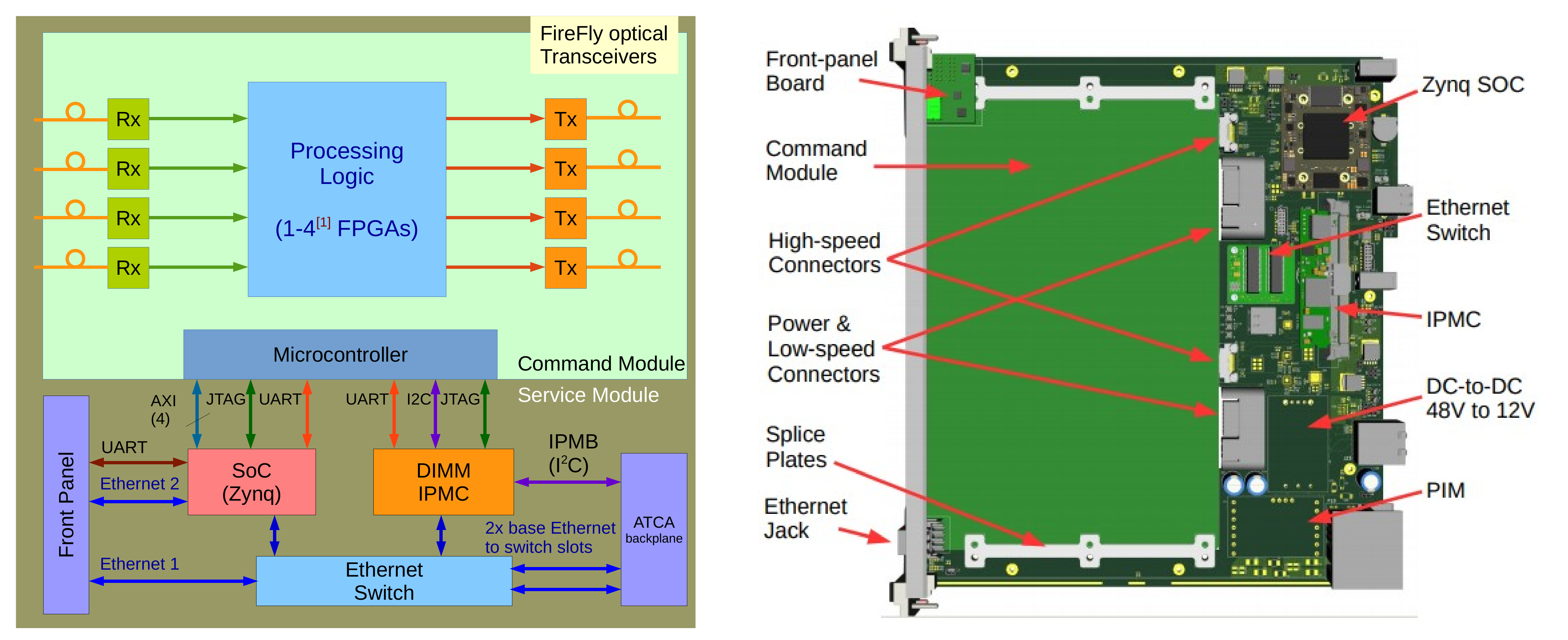}
    \caption{Apollo block diagram (left) and mechanics (right).}
    \label{fig:mech}
  \end{centering}
\end{figure}

The ATCA standard defines a front board with dimensions
8U~x~280~mm~x~6~HP equipped with a face plate, top and bottom handles,
indicator LEDs and a set of rear-facing backplane connectors.
The Apollo service module (SM) is an ATCA-compliant front board with a
large (7U~x~180~mm) cutout in which a command module (CM) containing
application-specific processing elements (see
Fig.~\ref{fig:mech}). The top surfaces of the SM and
CM PCBs are co-planar but the PCBs need not be of identical
thickness.  A minimum of 2 board-to-board connectors provide
electrical and mechanical connectivity between CM and SM, and the
boards are securely joined by metal splice plates.

A block diagram is shown in Fig.~\ref{fig:mech}.
The SM contains power, communications and control infrastructure,
providing \iic, UART, JTAG and AXI chip-to-chip links to the CM.
The CM can accommodate any hardware required for a particular
application but the demonstrator contains two FPGAs as required
by a typical LHC trigger/readout board.

\section{IPMC}

\label{sec:ipmc}

For the Apollo demonstrator we use the CERN IPMC\cite{bib:cern-ipmc},
which is based on a hardware and software core provided by Pigeon
Point.  The essential functions of the IPMC are: (1) to read a set of
critical parameters (temperature, current, voltage) from on-board
sensors and deliver them to the ATCA shelf manager and (2) to mange
power-up and power-down of the blade.

We have customized the IPMC extensively to perform several functions
not foreseen by the ATCA standard: Xilinx Virtual Cable server for
JTAG access; command interpreter accessible via TCP for GPIO and \iic
debugging; serial-over-LAN access to console ports of System on Chip
(SoC) and CM microcontroller; and a more sophisticated power-up and
power-down sequence.


\section{SoC and Ethernet}

A Xilinx Zynq SoC provides for overall control, monitoring, run
control, and local DAQ functions for the blade.  The demonstrator
contains an Enclustra Mercury ZX1 module with a XC7Z035/45, providing
a dual-core ARM Cortex-A9 CPU, 1~GB DDR3 SDRAM and two Ethernet ports.
A total of four high-speed links (up to 10~Gbps) are routed to the SM
for AXI chip-to-chip use.  In the production hardware, the SoC will be
upgraded to one supporting at least 64-bit CPUs and 4~GB of RAM.

Ethernet access is provided to the IPMC and SoC.  An unmanaged
Ethernet switch (ESM\cite{bib:esm}) has a total of 6
ports, of which 5 are used as follows:  2 for ATCA base
Ethernet from the backplane; one for a front-panel RJ-45 jack; one
for SoC access, and one for the IPMC.
An additional RJ-45 jack provides access to a dedicated gigabit
Ethernet port on the SoC.

\section{Clocks}
\label{sec:clocks}

\begin{figure}
  \includegraphics[width=\textwidth]{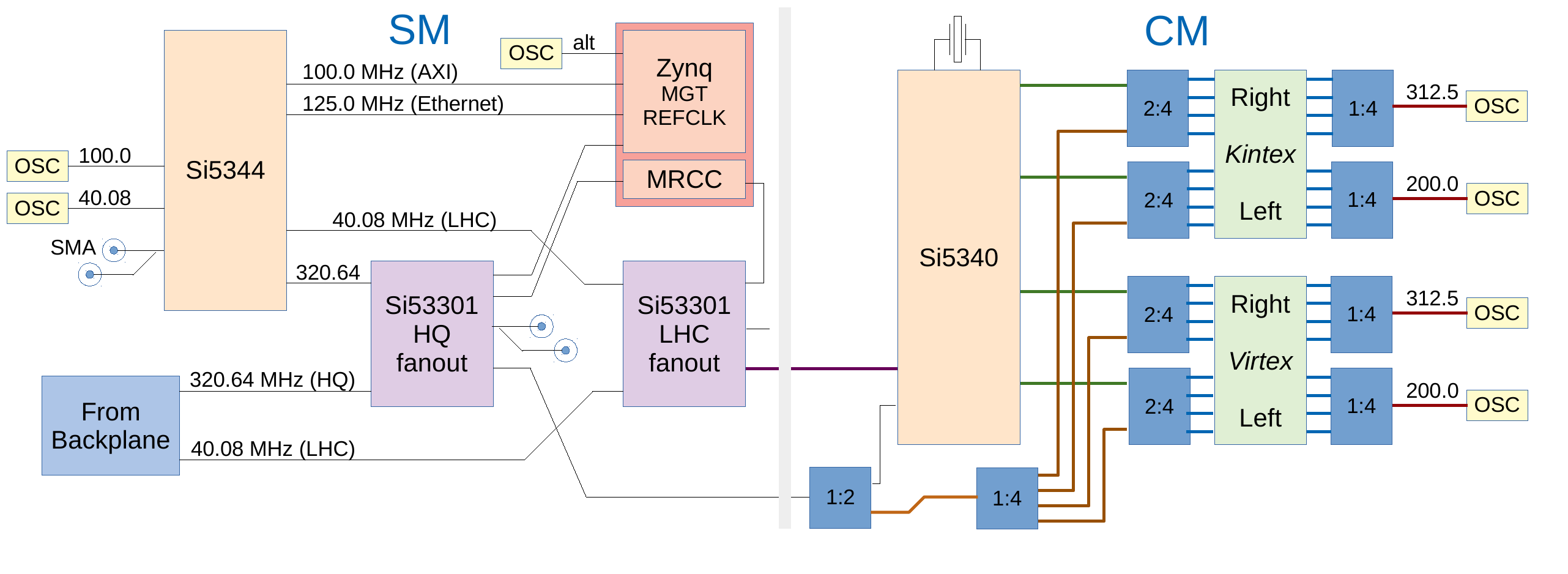}
  \caption{Apollo demonstrator clocking}
  \label{fig:clocks}
\end{figure}

An overview of Apollo demonstrator clock distribution is shown 
in Fig.~\ref{fig:clocks}.  The SM can
receive LHC machine clocks over the backplane or synthesize them
from crystal oscillators on-board.  The LHC clock is used
for optical links from the detector, driven by the CERN LpGBT
ASIC\cite{bib:lpgbt}, while fixed-frequency clocks are used for
Ethernet, AXI and 25~Gbps class optical links.

On the CM, clocks provided by the SM are distributed to a set of FPGA
reference clock inputs, while an additional set of inputs is driven by
local oscillators for bench testing.

\section{Application-Specific Processing}

The Apollo command-module is intended to be customized for each
application, but they share many common features such as large FPGAs
with ample logic resources and many high-speed fiber optic links.
Three specific applications and their anticipated hardware needs are
briefly presented.

The {\bf CMS Pixel DAQ and Timing Card} (DTC) receives data from up
to 512 front-end ASICs which deliver data in a complex, compressed
data format.  The total data rate for one DTC can approach 300~Gbps.
Triggers are received from the CMS central trigger and must be
converted to tokens and transmitted to the front-end. The received
compressed data stream must be parsed in real time to build events.  A
separate trigger stream is processed for CMS luminousity
monitoring system.  The DTC is currently foreseen to require two
XCVU7P or similar Virtex Ultrascale+ class FPGAs, 72 optical
links at 10~Gbps and 16 optical links at 25~Gbps.
A total of 28 pixel DTCs are required for CMS.

The {\bf CMS track finder} uses pattern-matching to identify coincidences
between ``tracklets'' transmitted from the readout modules and then
uses a Kalman filter to establish precise track parameters.
This application requires a smaller number of links than the DTC but
substantial FPGA resources.  The track finder is expected to require one
large or two smaller Virtex Ultrascale+ FPGAs and 60 optical
links at 25~Gbps.  A total of between 126 and 180 track finder blades
are required for CMS.

The {\bf ATLAS Monitored Drift-Tube Trigger Processor} (MDTTP)
performs a function similar to the combined functions of the DTC and
track finder for CMS, but for the muon drift-tubes in ATLAS.  Drift-tube 
hit data are received on about 60 fiber optic links, and a
sophisticated two-dimensional fit is used to identify track segments.
The segments are then joined to form tracks, and the Zynq processor is
used to calculate transverse momentum for the identified tracks.  In
addition, drift-tube hits are buffered and stored until a trigger is
received after which they are built into an event and sent to the DAQ.  
The MDTTP is expected to require one large Kintex Ultrascale+ FPGA
and one Zynq Ultrascale+ SoC, along with 72 optical links at 10~Gbps and
12 optical links at 25~Gbps.  A total of 64 MDTTP blades are
required for ATLAS.

\section{Prototypes and Test Results}

\begin{figure}
\begin{centering}
  \includegraphics[width=\textwidth]{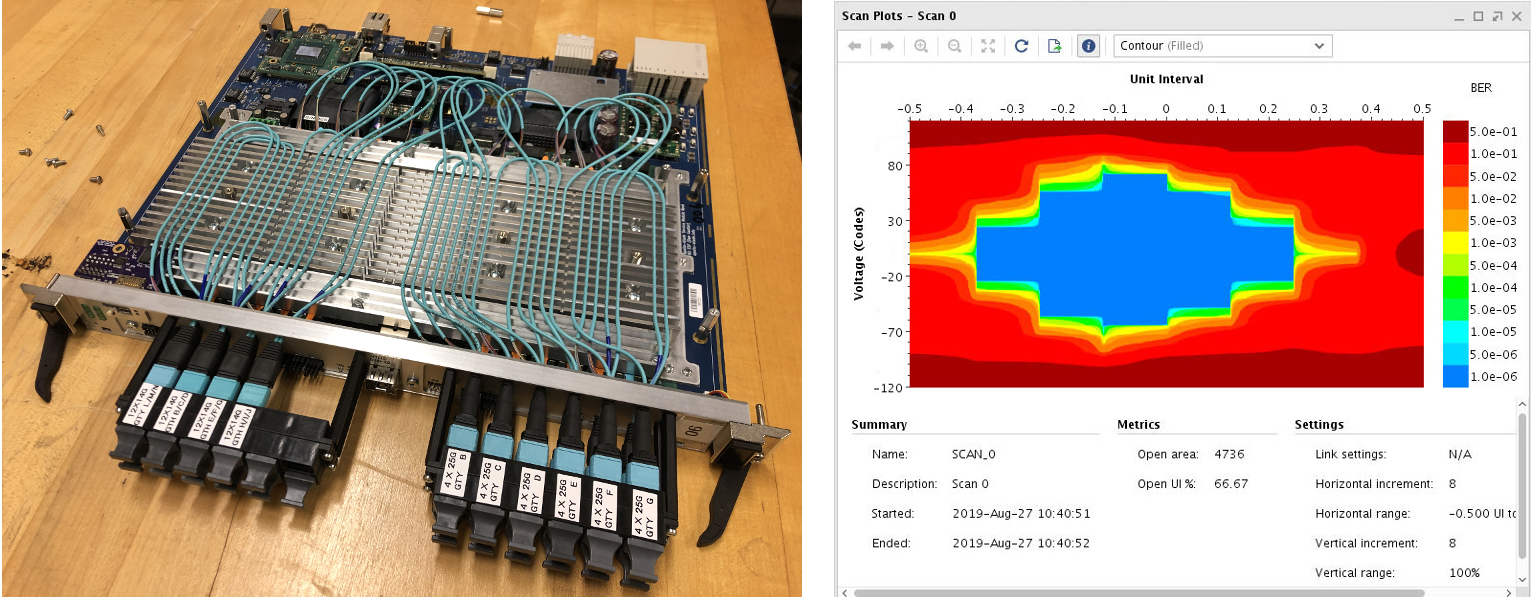}
  \caption{Demonstrator blade (left) -- 25~Gbps eye pattern (right)}
  \label{fig:demo}
\end{centering}  
\end{figure}

A series of 10 demonstrator boards (Fig.~\ref{fig:demo}) have been
constructed.  Each consists of a service module and a command module,
and can recieve two FPGAs, an XCKU15P in an A1760 package, and an
XCVU7P in a B2104 package.  A total of 88 FireFly channels
rated for 28~Gbps and 36 FireFly channels rated for 14~Gbps are
provided along with the flexible clocking and power architecture
described earlier.

Apollo blades have been deployed in several shelves at CERN and
collaborator institutes.  Remote access including firmware
reprogramming and remote reset/power cycle have been proven.  Optical
links have been tested extensively at 10~Gbps and 25~Gbps and shown to
be error-free.

The detailed design of customized command modules for specific
applications is underway.

A CM developed at Cornell University has been mated with a 
SM developed at Boston University, as shown in Fig.~\ref{fig:demo}.  
One KU15P and one VU7P Ultrascale+ FPGAs are installed.  Two large heat 
sinks cool the FPGAs while two smaller heat sinks cool the FireFly optical 
transceivers.  Optical fiber bundles are pressed into grooves machined 
in the heat sinks.

Figure~\ref{fig:demo} (right) shows an eye pattern for a fiber optic link
test with loop-back fiber between two different quads on the XCVU9P FPGA.
The horizontal axis is in fraction of one bit period (40~ps) and the vertical 
axis is arbitrary amplitude units.  This represents a good
result as logic ``0'' and logic ``1'' levels are well resolved in the
middle of the bit period.  Many combinations of links have been
operated for hours with no bit errors seen.

\section{Summary}

We have developed a novel Apollo ATCA blade, designed for a wide range
of applications.  It is foreseen to be used in the CMS Pixel readout, the CMS
track trigger, and the ATLAS drift-tube trigger processor. 


\begin{thebibliography}{99}

\bibitem{bib:esm}
  W.~Smith,
  \emph{Next Generation ATCA Control Infrastructure for the CMS
    Phase-2 Upgrades}, \\
  in TWEPP 2017,
  \pos{PoS(TWEPP-17)102}

\bibitem{bib:Serenity}
  A.~Rose,
  \emph{Serenity - An ATCA prototyping platform for CMS Phase-2}, \\
  in TWEPP 2018,
  \pos{PoS(TWEPP-18)115}

\bibitem{bib:cern-ipmc}
  J.~Mendez,
  \emph{CERN-IPMC Solution for AdvancedTCA Blades}, \\
  in TWEPP 2018, 
  \pos{PoS(TWEPP2018)059}

\bibitem{bib:lpgbt}
  P.~Moreira, 
  \emph{The lpGBT: a radiation tolerant ASIC for Data, Timing, Trigger
    and Control Applications in HL-LHC},
  in TWEPP 2019,
  \pos{PoS(TWEPP2019)}

  
\end{thebibliography}
\end{document}